# TotalSegmentator MRI: Robust Sequence-independent Segmentation of Multiple Anatomic Structures in MRI

Tugba Akinci D'Antonoli, MD[1] • Lucas K. Berger [1] • Ashraya K. Indrakanti, MD[1] • Nathan Vishwanathan, MD[1] • Jakob Weiss, MD[2] • Matthias Jung, MD[2,3] • Zeynep Berkarda, MD[2] • Alexander Rau, MD[2] • Marco Reisert, PhD[2] • Thomas Küstner, PhD[4] • Alexandra Walter, [5,6] • Elmar M. Merkle, MD[1] • Daniel Boll, MD[1] • Hanns-Christian Breit, MD[1] • Andrew Phillip Nicoli, MD[1] • Martin Segeroth, MD[1] • Joshy Cyriac, MSc[1] • Shan Yang, PhD[1] • Jakob Wasserthal, PhD[1]*

1 Clinic of Radiology and Nuclear Medicine, University Hospital Basel, Petersgraben 4,
CH-4031 Basel, Switzerland
2 Department of Diagnostic and Interventional Radiology, Faculty of Medicine, Medical Center - University of Freiburg, University of Freiburg
3 Department of Stereotactic and Functional Neurosurgery, Faculty of Medicine, Medical Center University of Freiburg, University of Freiburg, Freiburg, Germany
4 Medical Image and Data Analysis, University Hospital of Tuebingen, Tuebingen, Germany
5 Department of Biomedical Physics in Radiation Oncology, German Cancer Research Center, Heidelberg, Germany
6 Scientific Computing Center, Karlsruhe Institute of Technology, Karlsruhe, Germany

**Corresponding author**: jakob.wasserthal@usb.ch

# Abstract

**Background**

Since the introduction of TotalSegmentator CT, there is demand for a similar robust automated MRI segmentation tool that can be applied across all MRI sequences and anatomic structures.

**Purpose**

To develop and evaluate an automated MRI segmentation model for robust segmentation of major anatomic structures independently of MRI sequence.

**Materials and Methods**

In this retrospective study, a nnU-Net model (TotalSegmentator) was trained on MRI and CT examinations to segment 80 anatomic structures relevant for use cases such as organ volumetry, disease characterization, surgical planning and opportunistic screening. Examinations were randomly sampled from routine clinical studies to represent real-world examples. Dice scores were calculated between the predicted segmentations and expert radiologist reference standard segmentations to evaluate model performance on an internal test set, two external test sets and against two publicly available models, and TotalSegmentator CT. The model was applied to an internal dataset containing abdominal MRIs to investigate age-dependent volume changes. Wilcoxon signed-rank test was used.

**Results**

A total of 1143 examinations (616 MRIs, 527 CTs) (median age 61 years, IQR 50-72) were split into training (n=1088, CT and MRI) and an internal test set (n=55; only MRI), two external test sets (AMOS, n=20; CHAOS, n=20; only MRI), and an internal aging-study dataset of 8672 abdominal MRIs (median age 59 years, IQR 45-70) were included. The model showed a Dice Score of 0.839 on the internal test set and outperformed two other models (Dice Score, 0.862 versus 0.759; and 0.838 versus 0.560; p<.001 for both). On the TotalSegmentator CT test set (n=89), the proposed model almost matched the performance of TotalSegmentator CT (Dice Score, 0.966 versus 0.970; p<.001). The aging study demonstrated strong correlation between age and organ volumes (e.g., age and liver volume had rs: -0.096, p<.001).

**Conclusion**

The proposed open-source, easy-to-use model allows for automatic, robust segmentation of 80 structures, extending the capabilities of TotalSegmentator to MRIs of any sequence. The ready-to-use online tool is available at https://totalsegmentator.com, the model at https://github.com/wasserth/TotalSegmentator, and the dataset at https://zenodo.org/records/14710732.

# Introduction

MRI is an indispensable tool in diagnostic imaging, providing detailed images of the human body without the use of ionizing radiation essential for diagnosing a range of medical conditions, from neurological disorders to musculoskeletal injuries (1,2). While MRI offers unparalleled detail, the manual segmentation of these images requires intensive effort by radiologists, a time-consuming process, subject to interrater variability, and prone to errors.

The advent of automated image segmentation techniques has shown promise in addressing these limitations. One notable advancement in medical image segmentation is the development of nnU-Net, a self-configuring framework that has set new standards in medical image segmentation. It adapts to any new dataset with minimal user intervention, automatically adjusting its architecture, preprocessing, and training strategies to optimize performance (3,4). Tools such as TotalSegmentator explored the capabilities of nnU-Net so far on CT images and proved to be very robust and widely used by the medical imaging community (5). These automated systems can potentially reduce radiologist's workload, minimize human errors, and provide more consistent and reproducible results, with several clinical applications, such as treatment planning, monitoring disease progression, and opportunistic screening (6).

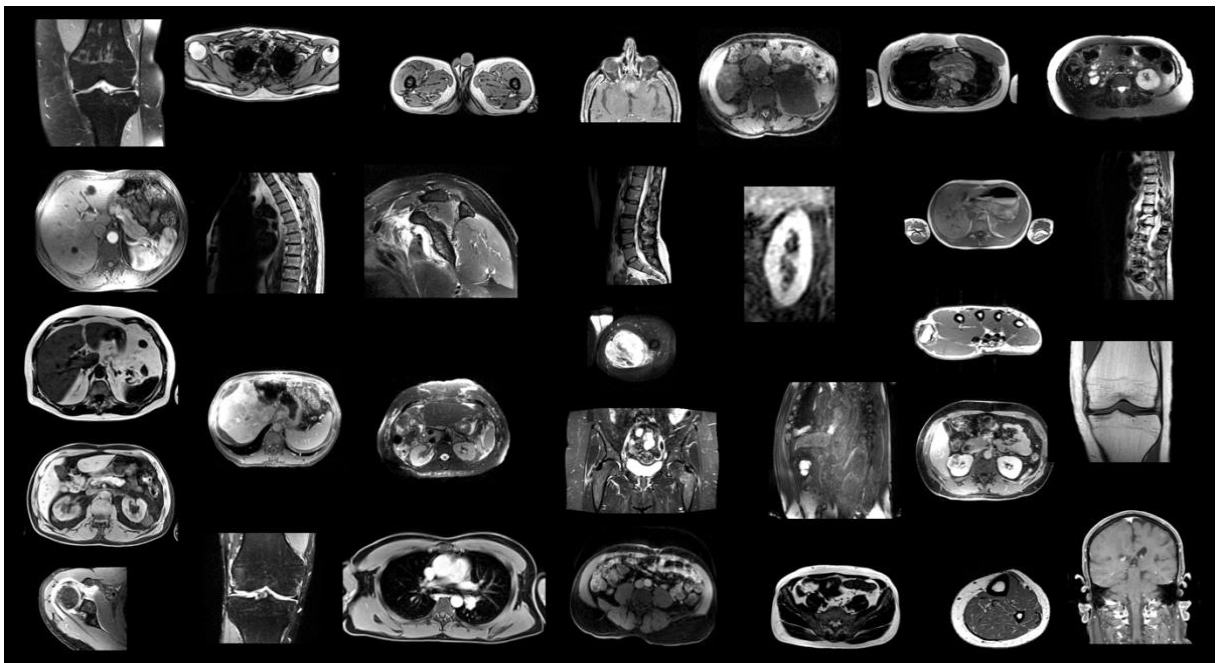

**Figure 1.** Exemplary overview of MRIs in the training dataset (n= 561). Since images were randomly sampled from the clinical routine it contains a wide variety of different contrasts, pathologies, and image types.

Despite these advancements in the field of segmentation of CT images, several challenges remain for the segmentation of MRIs. One issue is the variation in the MRI examinations due to different imaging parameters and protocols used across different sequences and body parts, which can affect the generalizability and accuracy of segmentation algorithms (7). Moreover, while existing algorithms perform well on some high-resolution sequences in which structures are well-defined with high contrast, they struggle with different sequence types and pathologically altered anatomic structures in which delineation is less clear. These factors highlight the need for ongoing research to enhance the robustness and reliability of automated MRI segmentation tools applicable across all sequences and all anatomic structures.

Thus, to extend the capabilities of TotalSegmentator to MRIs of any type, we aim to develop and evaluate an open-source, easy-to-use segmentation model for automatic and robust segmentation of major anatomic structures independent of MRI sequence.

# Materials and Methods

this retrospective study was approved by the Ethics Committee Northwest and Central Switzerland (EKNZ BASEC 2023-00446) and the requirement for informed consent was waived as this was a retrospective study.

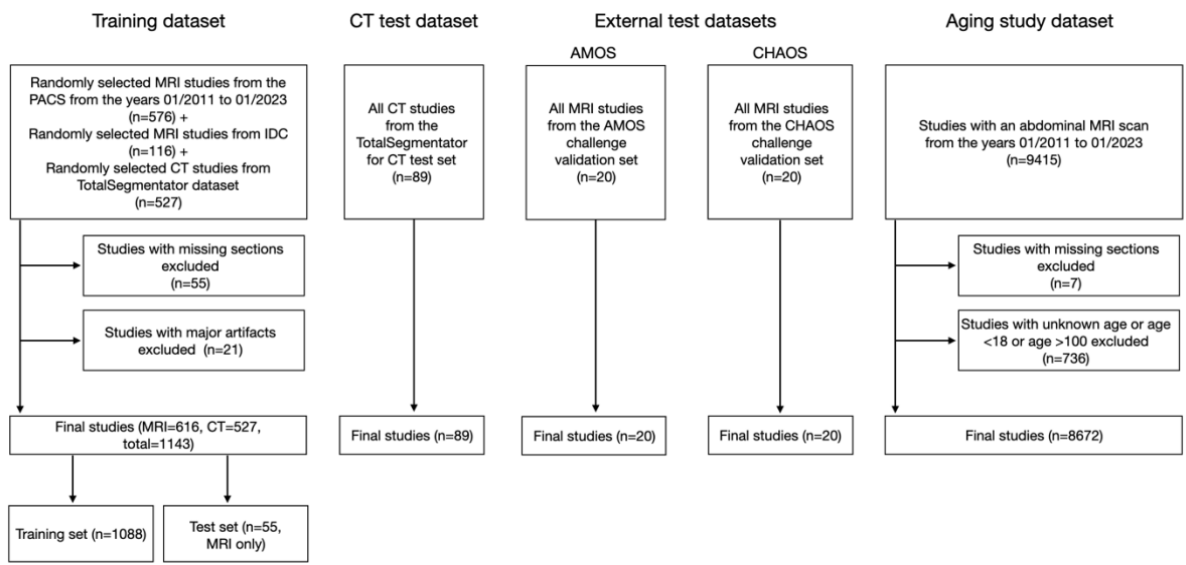

**Figure 2.** Diagram showing the inclusion of patients into the study. IDC = Imaging Data Commons, PACS = picture archiving and communication system, AMOS22 Grand challenge dataset available at: https://zenodo.org/records/7262581) and the CHAOS challenge dataset available at: http://doi.org/10.5281/zenodo.3362844.

## Datasets

### Training and internal test dataset

To generate a comprehensive and highly variant dataset, 576 MRI examinations were randomly sampled from the years 01/2011 to 01/2023 from the University Hospital Basel picture archiving and communication system. For each MRI examination, the series was sampled randomly to obtain a high variety of data with different kinds of MRI sequences (see **Figure 1** and results section for more details). Additionally, we added MRI examinations from Imaging Data Commons (8) to increase the image diversity (**Supplemental Material S1**. To make the model more robust a random set of CT examinations were added from the TotalSegmentator training dataset, to keep the dataset balanced with regards to the MRI examinations only about half of the TotalSegmentator training dataset was included (5). Images with minor artifacts, such as minimal flow artifacts, were included to enhance model robustness. Images with major artifacts like metal artifacts or clinically significant motion artifacts were excluded due to their potential to severely distort the image quality. Moreover, studies with missing sections and patients with unknown age or age <18 or >100 are excluded. For additional internal evaluation and direct comparison with TotalSegmentator CT model we used the test set from the TotalSegmentator paper, which we refer to as "CT test set".

Hyperparameter optimization was not performed. Instead, the proposed model used nnU-Net default values; thus, an additional validation set was not used. **Figure 2** provides a flow diagram of the inclusion of images in the datasets.

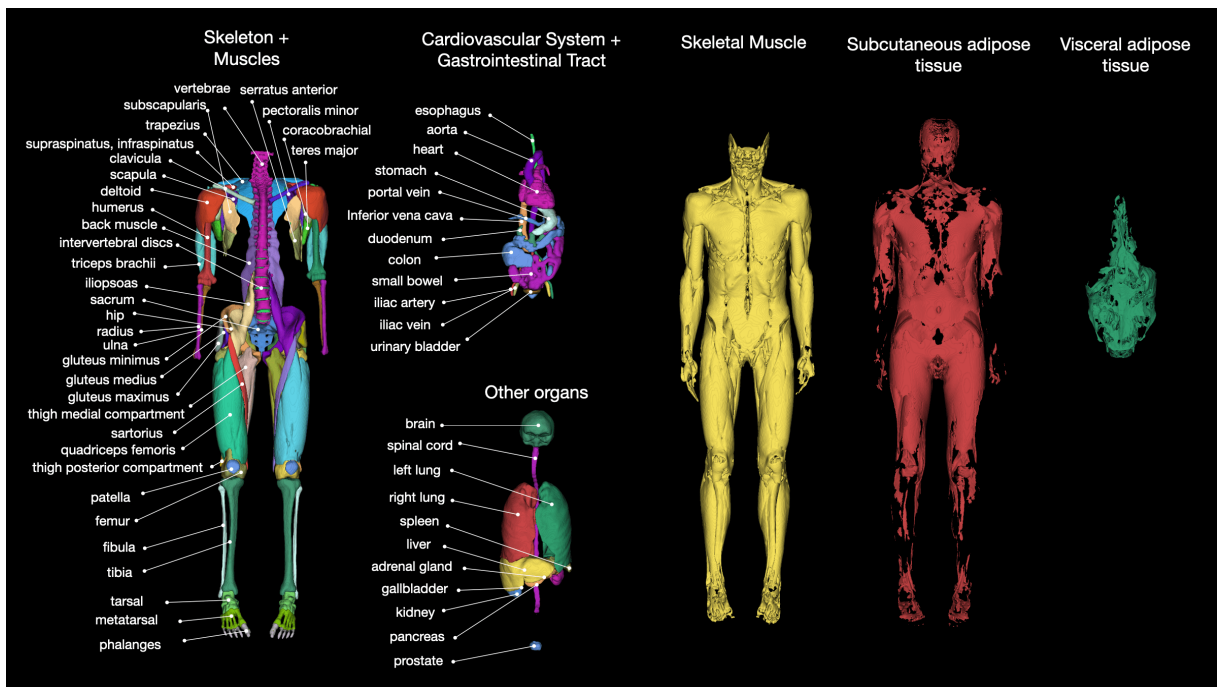

**Figure 3.** Overview of supported anatomic structures

### External test datasets

For external test sets, 20 MRI examinations from the AMOS22 challenge test set (9) (dataset available at: https://zenodo.org/records/7262581) and 20 MRI examinations from the CHAOS challenge test set (10) (dataset available at: http://doi.org/10.5281/zenodo.3362844) were used.

### Internal Aging-study dataset

All patients who received an abdominal MRI examination with T1-weighted sequence between 01/2011 and 01/2023 at the University Hospital Basel were consecutively included (n = 9408). Patients with unknown age (n = 80), age below 18 (n = 652) or above 100 (n = 4) were excluded. Moreover, studies with missing sections are excluded (n = 7) (**Figure 2**).

## Data Annotation

80 anatomic structures were identified for segmentation (**Figure 3**; **Supplemental Material S2**). The Nora Imaging Platform was used for manual segmentation or further refinement of generated segmentations (11). The initial segmentations were performed independently by two board-certified radiologists (TAD with 12 years of experience and HCB with 7 years of experience), who also supervised the iterative learning process. If an existing model for a given structure was publicly available (**Supplemental Material S3**), that model was used to create the first segmentation, which was then reviewed and refined manually (12,13).

To speed the process further, an iterative learning approach was used as previously described (5). Namely, after manual segmentation of the first 10 patients was completed, a preliminary nnU-Net was trained, and its predictions were manually refined, if necessary. Retraining of the nnU-Net was performed after reviewing and refining 125 additional patients. In the end, all 616 MRI examinations had annotations that were manually reviewed and corrected if necessary. These final annotations served as the reference standard for training and testing. This final model was independent of the intermediate models trained during the annotation workflow, which reduced bias in the test set to a minimum.

Since the two external test sets (AMOS and CHAOS) do not contain reference segmentation masks for all anatomic structures, the missing ones were manually added by a board-certified radiologist (TAD with 12 years of experience) using the Nora Imaging Platform.

All annotators were blinded to clinical information, as they had access only to the images provided on the annotation platform, without any access to radiology reports.

## Model

A model from the nnU-Net framework, a U-Net–based implementation, was utilized. This framework automatically configures all hyperparameters based on the dataset characteristics (3,4). One model was trained on MRI+CT images with 1.5 mm isotropic resolution (baseline resolution of the training dataset), i.e., TotalSegmentator MRI. To allow for lower technical requirements (random-access memory [RAM] and graphics processing unit [GPU] memory), a second model was also trained on 3 mm isotropic resolution, i.e., TotalSegmentator MRI-3 (for more details on the training see **Supplemental Material S4**).

Training one model with 1.5 mm resolution and all 80 anatomic structures, namely TotalSegmentator MRI, would require more RAM than typical workstations have available. Therefore, the anatomic structures were split into 6 models. For the comparison to the TotalSegmentator MRI-3, only 50 of the most important anatomic structures were selected (see **Supplemental Material S5**). Otherwise, the required training time for all model combinations would have exceeded our available resources.

The runtime for the prediction of one case was measured on a local workstation with an Intel Core i9 3.5GHz central processing unit and Nvidia GeForce RTX 3090 GPU.

## Baseline Models

TotalSegmentator MRI was compared against the following two publicly available baseline models: 1) The tool MRSegmentator (13) and 2) a nnU-Net ("AMOS") that trained on the AMOS22 Grand Challenge training set. (9). The comparison was limited to 40 structures supported by MRSegmentator and 13 structures supported by AMOS (**Supplemental Material S3**).
Moreover, the proposed model was compared against TotalSegmentator CT.

# Ablation study

For the ablation study, two additional models were trained: one exclusively on MRIs and another exclusively on CT images from our training dataset. Their performances were compared to that of the TotalSegmentator MRI-3 model for the 50 most important structures (**Supplemental Material S5**).

Both the annotated training dataset (https://zenodo.org/doi/10.5281/zenodo.11367004) and the code for the open-source proposed model (https://www.github.com/wasserth/TotalSegmentator) are publicly available. Moreover, a ready-to-use online tool is also available at https://totalsegmentator.com.

# Statistical Analysis

## Training dataset

As evaluation metrics, the Dice similarity coefficient, also known as the Dice score, a commonly used spatial overlap index, and the normalized surface distance (NSD), which measures how often the surface distance is <3 mm, were calculated between the predicted segmentations and the human-approved reference standard segmentations as recommended (14). Both metrics range between 0 (worst) and 1 (best) and were calculated on the test set. For additional evaluation, we compared our model also on the test set from the original TotalSegmentator paper containing only CT images (5).

Normal distribution of the Dice Score and NSD was rejected by using the Kolmogorov-Smirnov test. Continuous variables were reported as pseudomedian and associated 95% CIs using an underlying signed rank distribution. Wilcoxon signed-rank test was used for model comparison with a *P* value of less than .05 considered statistically significant. All statistical analyses were performed using Python 3.9.

## Aging study dataset

The correlation between age and organ volume changes was calculated using the following abdominal structures, left and right adrenal gland, left and right kidney, liver, and spleen, to evaluate the effects of aging. The volume was calculated using the formula: *volume = number of voxels × voxel volume*. The segmentation of a structure was assumed as failed and excluded if the volume of the respective body structure was too small to be anatomically plausible given by a lower bound (**Supplemental Table**). The Kolmogorov-Smirnov test was used to evaluate whether continuous variables were normally distributed. The association between continuous variables was examined using Spearman's rank correlation coefficient. Patients were grouped into four age quartiles and compared using the Kruskal-Wallis test. Post-hoc analysis was performed using the Wilcoxon rank sum test. Bonferroni correction was performed and p-values of <0.0001 were considered significant. Outliers are not shown in the figures to maintain scaling. All statistical analyses were performed using Python 3.9.

# Results

## Dataset Characteristics

A total of 1143 examinations (616 MRI, 527 CT) were included in this study and split into training (n = 1088; CT and MRI) and an internal test set (n = 55; only MRI) at the patient level. The MRI training dataset contains 258 (51.6%) males and 242 (48.4%) females and the median age is 61.0 (IQR 50.0-72.0) (see Figure 2 and Table 1). The dataset contained a wide variety of MRI examinations, with differences in contrast, section thickness, field strength, pulse sequences (e.g. T1-weighted, T2-weighted, Proton Density), echo time, repetition time, resolution, and contrast agent. The MRIs were acquired by 30 different scanners from three different manufacturers and at 4 different sites. The distribution of the MRI dataset is shown in **Figure 4**. The additional 527 CTs were randomly sampled from the TotalSegmentator training dataset and followed the distributions shown in Figure 3 of the TotalSegmentator CT paper (5). Additionally, we added 116 MRIs from Imaging Data Commons (8) to increase the image diversity (S4).

For additional internal evaluation, "CT test set" consisting of 89 CTs from the TotalSegmentator CT test set (4) was used. Moreover, two external test sets (AMOS, n = 20; CHAOS, n = 20) containing sole MRIs were used.

The aging-study dataset consisted of MRIs only from 8672 patients with an age range 18-100 years (median 56 years, IQR 45-70) (**Figure S1**) and a sex distribution of 4454 (51.4%) males and 4218 (48.6%) females (**Table 1**).

Final dataset characteristics can be found in **Figure 4.**

**Table 1.** Demographic characteristics of the datasets

A: Training dataset

|  | **MRI dataset (n = 616)** | **CT dataset (n = 527)** |
|---|---|---|
| **Age range (years)** | **18 - 100** | **18 - 100** |
| **Median age (years)** | 61.0 (IQR: 50.0-72.0) | 70.0 (IQR: 60.0-80.0) |
| **Sex** | | |
| ● Male | 258 | 247 |
| ● Female | 242 | 184 |
| ● Unknown | 116 | 96 |

**B.** Internal aging study dataset

|  | **MRI dataset (n = 8672)** |
|---|---|
| **Age range (years)** | **18 - 100** |
| **Median age (years)** | 59.0 (IQR: 45.0-70.0) |
| **Sex** | |
| ● Male | 4454 |
| ● Female | 4218 |

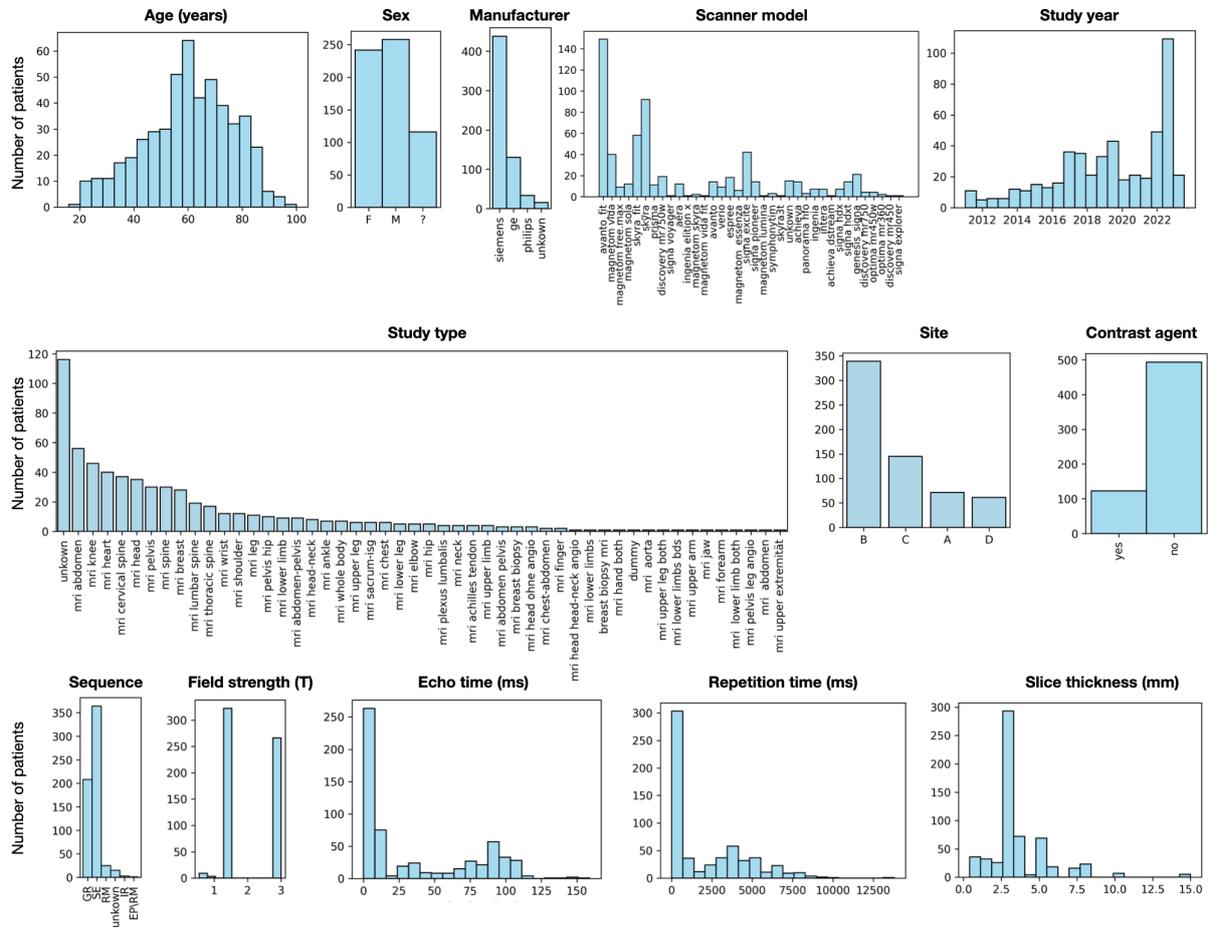

**Figure 4.** Histograms showing the distribution of different parameters of the training dataset, demonstrating the dataset's high diversity. (n = 561 MRIs and 527 CTs).

## Segmentation Evaluation

The proposed model (TotalSegmentator MRI) showed a Dice score of 0.839 [95% CI: 0.825, 0.851], an NSD of 0.907 [95% CI: 0.895, 0.919] for all 80 structures. For the 50 most important structures (**Supplemental Material S5**) proposed model showed a Dice score of 0.862 [95% CI: 0.849, 0.874], an NSD of 0.941 [95% CI: 0.931, 0.950] and the TotalSegmentator MRI-3 model showed a lower performance with a Dice score of 0.779 [95% CI: 0.757, 0.799], and a NSD of 0.858 [95% CI: 0.839, 0.877] (p<0.001). Results for each structure independently are shown in **Figure S2** and at https://github.com/wasserth/TotalSegmentator/blob/master/resources/results_all_anatomic_structures_mr.json.

Compared with MRSegmentator, a publicly available model, the proposed model achieved a higher Dice score (0.862 [95% CI: 0.846, 0.875] vs 0.759 [95% CI: 0.734, 0.780]); p<0.001) and NSD score (0.932 [95% CI: 0.920, 0.943] vs 0.833 [95% CI: 0.806, 0.857]; p<0.001) on the 40 structures supported by MRSegmentator.

The proposed model demonstrated better performance than the AMOS model, with a Dice score of 0.838 [95% CI: 0.802, 0.864] compared to AMOS's 0.560 [95% CI: 0.502, 0.618], and an NSD of 0.919 [95% CI: 0.890, 0.940] versus AMOS's 0.655 [95% CI: 0.598, 0.706]; both comparisons yielded p<0.001 on the 13 abdominal structures supported by AMOS model.

On the external AMOS dataset, the model performed better than MRSegmentator (Dice score of 0.886 [95% CI: 0.873, 0.898] vs 0.860 [95% CI: 0.847, 0.873]; p<0.001 and NSD of 0.946 [95% CI: 0.936, 0.954] vs 0.930 [95% CI: 0.920, 0.939]; p=0.002). The AMOS model showed the best performance with a Dice score of 0.907 [95% CI: 0.893, 0.919] and an NSD of 0.972 [95% CI: 0.965, 0.978], which might be explained by a high similarity between the AMOS training and test datasets, although its performance does not generalize to other datasets as shown in our analysis (Table 2).

On the external CHAOS test set, our model performed better than MRSegmentator (Dice score of 0.890 [95% CI: 0.881, 0.898] vs 0.819 [95% CI: 0.801, 0.834]; NSD of 0.976 [95% CI: 0.970, 0.980] vs 0.948 [95% CI: 0.940, 0.956]; both $P$ <0.001). The AMOS model showed the worst performance with a Dice score of 0.730 [95% CI: 0.694, 0.760] and an NSD of 0.897 [95% CI: 0.872, 0.914]. **Figure 5** shows qualitative results for the liver and spleen of the cases with the highest and lowest Dice score in the CHAOS dataset for all models.

On the CT test set, the TotalSegmentator CT performed best (Dice score: 0.970 [CI: 0.969, 0.971], NSD: 0.997 [95% CI: 0.996, 0.997]), followed by the proposed model, TotalSegmentator MRI (Dice score: 0.966 [95% CI: 0.965, 0.967], NSD: 0.996 [95% CI: 0.995, 0.996]; $P$<.001). This was followed by MRSegmentator (Dice score: 0.944 [95% CI: 0.942, 0.946], NSD: 0.984 [95% CI: 0.982, 0.985], $P$ <.001) and then AMOS (Dice score: 0.907 [95% CI: 0.901, 0.913], NSD: 0.961 [95% CI: 0.956, 0.965], $P$ <.001).

**Table 2** gives an overview of all results.

**Table 2.** Overview of results on the proposed nnU-Net segmentation model (TotalSegmentator MRI) and the baseline models on the MRI and the CT test set;

| Model | Number of anatomic structures | Dice score | NSD |
|---|---|---|---|
| **MRI test set** | | | |
| TotalSegmentator MRI* | 80 | 0.839 [0.825, 0.851] | 0.907 [0.895, 0.919] |
| TotalSegmentator MRI | 50 | 0.862 [0.849, 0.874] | 0.941 [0.931, 0.950] |
| TotalSegmentator MRI-3 | 50 | 0.779 [0.757, 0.799] | 0.858 [0.839, 0.877] |
| TotalSegmentator MRI | 40 | 0.862 [0.846, 0.875] | 0.932 [0.920, 0.943] |
| MRSegmentator | 40 | 0.759 [0.734, 0.780] | 0.833 [0.806, 0.857] |
| TotalSegmentator MRI | 13 | 0.838 [0.802, 0.864] | 0.919 [0.890, 0.940] |
| AMOS | 13 | 0.560 [0.502, 0.618] | 0.655 [0.598, 0.706] |
| **CT test set** | | | |
| TotalSegmentator MRI | 40 | 0.966 [0.965, 0.967] | 0.996 [0.995, 0.996] |
| TotalSegmentator CT | 40 | 0.970 [0.969, 0.971] | 0.997 [0.996, 0.997] |
| MRSegmentator | 40 | 0.944 [0.942, 0.946] | 0.984 [0.982, 0.985] |
| TotalSegmentator MRI | 13 | 0.956 [0.954, 0.959] | 0.994 [0.992, 0.995] |
| AMOS | 13 | 0.907 [0.901, 0.913] | 0.961 [0.956, 0.965] |
| **AMOS external test set (MRI only)** | | | |
| TotalSegmentator MRI | 40 | 0.886 [0.873, 0.898] | 0.946 [0.936, 0.954] |
| MRSegmentator | 40 | 0.860 [0.847, 0.873] | 0.930 [0.920, 0.939] |
| TotalSegmentator MRI | 13 | 0.807 [0.786, 0.830] | 0.886 [0.872, 0.900] |
| AMOS | 13 | 0.907 [0.893, 0.919] | 0.972 [0.965, 0.978] |
| **CHAOS external test set (MRI only)** | | | |
| TotalSegmentator MRI | 40 | 0.890 [0.881, 0.898] | 0.976 [0.970, 0.980] |
| MRSegmentator | 40 | 0.819 [0.801, 0.834] | 0.948 [0.940, 0.956] |
| TotalSegmentator MRI | 13 | 0.865 [0.851, 0.876] | 0.965 [0.957, 0.972] |
| AMOS | 13 | 0.730 [0.694, 0.760] | 0.897 [0.872, 0.914] |

Note—Data in brackets are 95% CIs. *The proposed model was trained on both MRI and CT images with their baseline resolution. The Dice score measures the degree of spatial overlap as 0 (no overlap) and 1 (perfect overlap), and the normalized surface distance (NSD) measures how often the surface distance is <3 mm. The proposed model, TotalSegmentator MRI, was compared against TotalSegmentator MRI-3, which contained only the 50 most important anatomic structures. The proposed model was also compared against the following publicly available baseline models: MRSegmentator, TotalSegmentator CT, and a nnU-Net (AMOS) trained on the AMOS22 Grand Challenge training dataset. See **Supplemental Material S2 and S4** for the list of segmented structures.

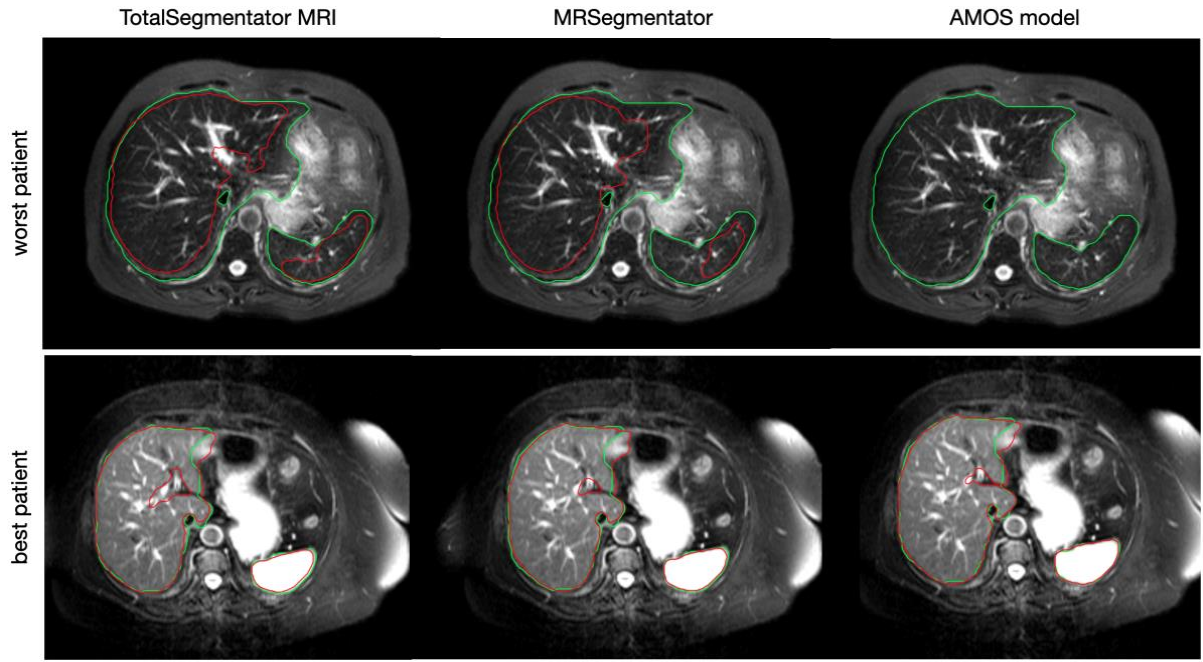

**Figure 5.** Patient with the best and worst Dice score in the CHAOS external test set for our proposed model, TotalSegmentator MRI, as well as for the two publicly available baseline models, MRSegmentator and AMOS. The reference segmentation for liver and spleen can be seen in green and the segmentation of the respective model in red. The CHAOS dataset was used to show the best and the worst results since this dataset is most independent from training data of the three models. CHAOS challenge dataset available at: http://doi.org/10.5281/zenodo.3362844.

## Ablation study

We trained two additional models: one solely on MRIs and another solely on CT images from our training dataset. Our proposed model, which was trained on both MRI and CT images, demonstrated better performance on the MRI test set compared to the model trained exclusively on MRIs (Dice score of 0.862 vs. 0.845 and NSD of 0.941 vs. 0.927, respectively, $p<.001$), which indicates that incorporating CTs enhances MRI segmentation. Conversely, on the CT test set, the model trained exclusively on CT images outperformed the combined MRI+CT model (Dice score of and NSD of 0.971 vs. 0.966, respectively, $p<.001$), which suggests that focusing on standardized Hounsfield unit values in CTs aids the model, while the inclusion of MRIs complicates this learning process. The complete results of the ablation study are presented in **Table 3**.

**Table 3.** Overview of ablation study results

| Model | Number of anatomic structures | Dice score | NSD |
| --- | --- | --- | --- |
| **MRI test set** | | | |
| trained on MRI + CT | 50 | 0.862 [0.849, 0.874] | 0.941 [0.931, 0.950] |
| trained on MRI | 50 | 0.845 [0.832, 0.857] | 0.927 [0.915, 0.936] |
| trained on CT | 50 | 0.040 [0.016, 0.075] | 0.075 [0.039, 0.108] |
| **CT test set** | | | |
| trained on MRI + CT | 50 | 0.966 [0.965, 0.967] | 0.996 [0.995, 0.996] |
| trained on MRI | 50 | 0.739 [0.729, 0.748] | 0.750 [0.740, 0.759] |
| trained on CT | 50 | 0.971 [0.970, 0.972] | 0.997 [0.996, 0.997] |

Note—Data in brackets are 95% CIs. The Dice score measures the degree of spatial overlap as 0 (no overlap) and 1 (perfect overlap), and the normalized surface distance (NSD) measures how often the surface distance is <3 mm. The 50 most important segmented structures for each model were included (**Supplemental Material S4**).

## Failure cases

In the MRI test set, lower Dice scores were observed compared to the CT test set. This discrepancy is primarily due to the inferior image quality of MRIs. Many MRIs exhibit high anisotropy, with section thicknesses exceeding 6 mm, making it challenging to detect small structures such as the iliac arteries and veins, sometimes even for the human eye. Additionally, other MRIs display low contrast outside the area of interest, further complicating the identification of structures in these regions. **Figure 6** shows some examples.

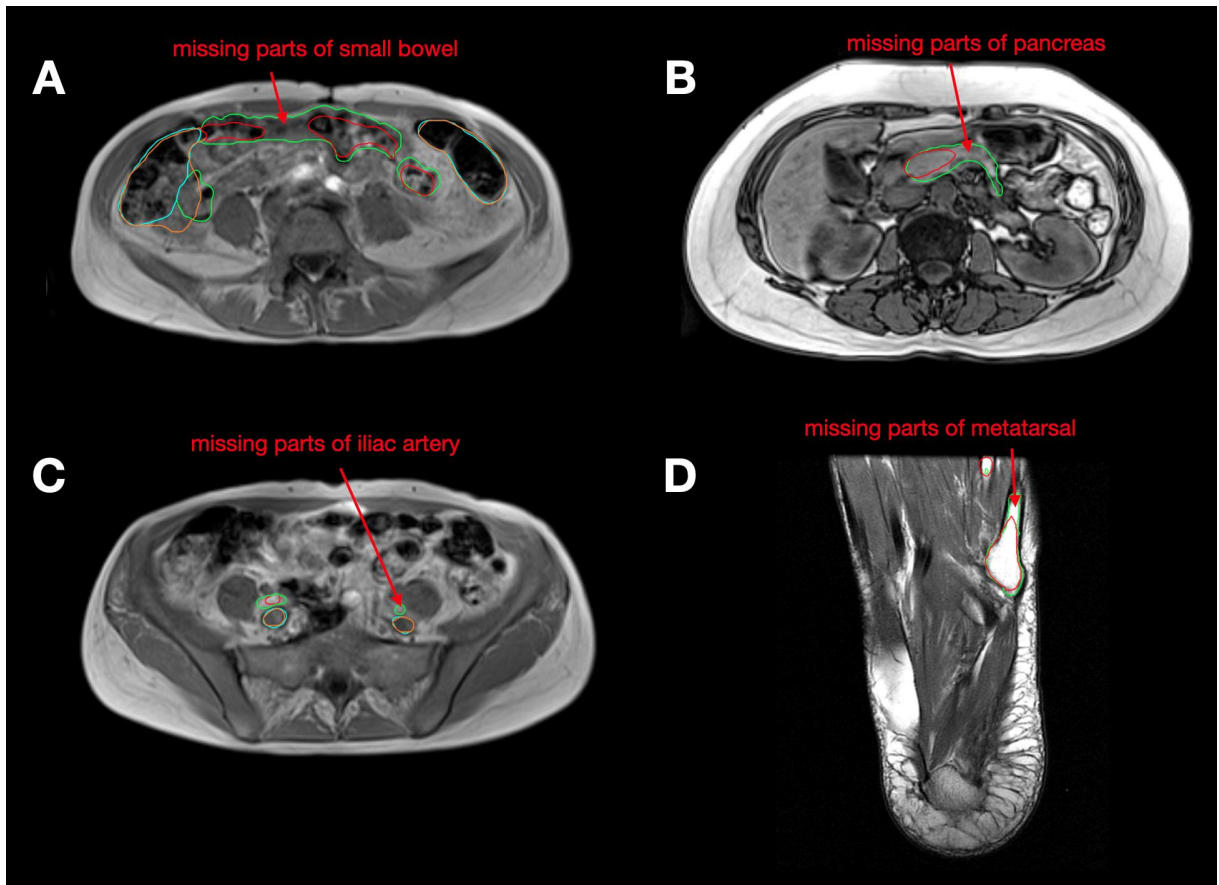

**Figure 6.** Examples of failure cases from the MRI test set. A: The small bowel model prediction (red) is missing parts in contrast to the reference segmentation (green). The colon prediction (orange) is over segmenting in contrast to the reference segmentation (cyan) B: The pancreas model prediction (red) is missing parts in contrast to the reference segmentation (green). C: The iliac artery prediction (red) is missing parts in contrast to the reference segmentation (green). The iliac vein prediction (orange) is very similar to the reference segmentation (cyan). D: The metatarsal model prediction (red) is missing parts in contrast to the reference segmentation (green).

# Runtime

Since our model uses the same nnU-Net architecture and the same baseline resolution of 1.5 mm and 3 mm as TotalSegmentator CT, for runtime and memory requirements, we refer to the previous paper (5).

# Evaluation of Age-related Differences

For the aging study dataset, we observed a positive correlation between age and volume of the left (rs = 0.323; p<0.0001) and right (rs = 0.310; p<0.0001) adrenal gland. A negative correlation was observed between patient age and the volume of the left (rs = −0.152; p<0.0001) and right (rs = −0.131; p<0.0001) kidney, the liver (rs = −0.096; p<0.0001) and the spleen (rs = −0.067; p<0.0001). **Figure S3** shows the distribution of volumes over age in more detail.

# Discussion

There is a need for robust automated MRI segmentation tools that can be applied across all MRI sequences and anatomic structures. In this study, we present the TotalSegmentator MRI, a robust model capable of automatically segmenting 80 anatomic structures with high accuracy, trained on a diverse dataset of MRIs and CTs of different anatomic regions acquired with different techniques. The proposed open-source, easy-to-use model allows for automatic, robust segmentation of MRIs, independent of sequence. and achieved good performance across 80 structures with a Dice score of 0.839. It significantly outperformed two other publicly available models (Dice score of 0.862 for proposed model vs. 0.759 MRSegmentator and 0.838 for proposed model vs. 0.560 AMOS). We have made our model, training dataset, and annotations openly available.

Since the introduction of TotalSegmentator CT (5), the demand for a robust model capable of automatically segmenting multiple structures in MRIs has increased (6). Following the similar iterative workflow that we had previously introduced with the TotalSegmentator CT, we developed the TotalSegmentator MRI. Compared to TotalSegmentator CT (5) on CT test set, both TotalSegmentator MRI and CT models achieved high Dice scores over 0.9 (0.966 vs 0.970, respectively; $P$<.001), reflecting modality-independent capabilities of TotalSegmentator MRI. While TotalSegmentator CT can automatically segment more structures, our current model can segment the highest number of structures on any MRIs among those specifically designed for MRI segmentation

As demonstrated in our ablation study, our proposed model, which was trained on both MRIs and CTs, outperformed the model trained exclusively on MRIs when evaluated on the MRI test set. This finding indicates that incorporating CT images enhances the model's performance for MRI segmentation and effectively serves as a form of data augmentation, a common practice in deep learning to improve model robustness.

Recently, new studies on automatic MRI segmentation have been published, but most have not shared their model and data (13,15,16). To our knowledge, our model is the only one that can automatically segment the highest number of structures on MRIs of any sequence.

Among these recent studies, only one study has made their model, MRSegmentator, available, but the authors did not share their training data or annotations (13). The authors trained the MRSegmentator, an nnU-Net model, in which the initial segmentations are generated by running TotalSegmentator CT on their training data, which were later confirmed (or corrected if necessary) by a radiologist, and later retrained the model based on these updated segmentations (13). Their model automatically segments 40 structures, and when we tested MRSegmentator on our training set, it achieved a median Dice score of 0.759 (versus Dice score 0.862 for TotalSegmentator MRI) on MRIs and achieved a median Dice score of 0.944 (versus Dice score 0.966 for TotalSegmentator MRI) on CTs for those 40 anatomic structures. Our TotalSegmentator MRI model offers a clear advantage over this new model as it can segment more structures and was trained on a more diverse clinical dataset. In addition, we have made both our dataset and annotations publicly available, unlike previous state-of-the-art models.

In another recent study, authors trained an nnU-Net model called MRAnnotator, which can segment 49 structures on MRIs (15). Authors reported that the MRAnnotator achieved a total average Dice score of 0.814 on the external dataset. Another recently published study introduced MRISegmentator, an nnU-Net model, which could segment 62 structures only on T1-weighted abdominal MRIs (16). The authors reported that MRISegmentator achieved an average Dice score of 0.829±0.133 (Mean and SD) on the AMOS22 dataset. Since these aforementioned models or their training datasets were not publicly available (15,16), we were unable to make a direct comparison and validate their results. These recent publications demonstrate the growing interest and further potential of advances in the field of medical image segmentation, particularly for MRIs, which could facilitate volumetric calculations of anatomic structures and support opportunistic screening. However, the lack of publicly available models and datasets significantly limits the benefits, as it hinders the ability of others to build upon or conduct meaningful comparisons.

Recent advancements in foundation models, such as the Segment Anything Model (SAM) (17) and its derivatives like MedSAM2 model (18), are promising; however, these models still require training data and cannot operate fully automated, particularly with 3D images. In practice, annotators must still manually draw bounding boxes around organs on each 2D images. In contrast, our proposed model is fully automated and requires no manual annotation, making a direct comparison with foundation models impractical. Moreover, a recent study (19) demonstrated that nnU-Net continues to outperform other segmentation models, including transformer-based ones, in 3D medical image segmentation when evaluated fairly across multiple datasets. This suggests that nnU-Net remains the leading model in this domain and would likely to outperform SAM as well as MedSAM2 in 3D segmentation until these newer models provide sufficient evidence.

In our aging study, we illustrated a clinical application of our segmentation model, demonstrating its potential to provide valuable insights into the age-related variations in organ volumes. Previously, such large-scale evaluations were either unfeasible or required substantial time and expertise. Using an internal dataset of over 8000 patients who underwent an abdominal MRI examination, we identified correlations between the age and volume of various organs. So far, the normal organ sizes and age-related organ development are typically derived from studies with small sample size in the literature. The potential applications of this model are extensive and include, but are not limited to, the assessment of organ size based on factors such as age, sex, or disease state.

Our study had several limitations. Our proposed model showed a lower performance on MRIs than on CT images, as MRI inherently display greater heterogeneity than CT images due to variations in acquisition protocols and scanner configurations, whereas CT images tend to be more homogeneous and generally have higher spatial resolution and signal-to-noise ratio. To mitigate this challenge, we used a diverse MRI training dataset, enabling consistent performance across the wide range of variability found in MRIs. Secondly, the retrospective design of our study represents a limitation. To mitigate this, we incorporated two external datasets; however, a prospective evaluation will be necessary in future research to further strengthen the findings.

In conclusion, we have developed an open-source, easy-to-use model for automatic and robust segmentation of MRIs using a highly diverse MRI training set. The proposed MRI segmentation tool extends the capabilities of TotalSegmentator to MRIs for most major anatomic structures, independent of MRI sequence. The ready-to-use online tool is available at [https://totalsegmentator.com](https://totalsegmentator.com). Our model can be easily integrated into existing clinical workflows and operate in real-time to assist radiologists during diagnostic processes. It can also be employed in various research projects, for example, analyzing age-dependent changes in the volume of different abdominal structures. In the future, we plan to enhance our model by adding more MRI structures for evaluation and expanding our training set.

**References**

1. Pooley RA. AAPM/RSNA physics tutorial for residents: fundamental physics of MR imaging. Radiographics. 2005 Aug;25(4):1087–99. DOI: 10.1148/rg.254055027. PMID: 16009826.
2. How does MRI work? Berlin, Heidelberg: Springer Berlin Heidelberg; 2006. DOI: 10.1007/978-3-540-37845-7.
3. Isensee F, Jaeger PF, Kohl SAA, Petersen J, Maier-Hein KH. nnU-Net: a self-configuring method for deep learning-based biomedical image segmentation. Nat Methods. 2021 Feb;18(2):203–11. DOI: 10.1038/s41592-020-01008-z. PMID: 33288961.
4. Isensee F, Petersen J, Klein A, Zimmerer D, Jaeger PF, Kohl S, et al. nnU-Net: Self-adapting Framework for U-Net-Based Medical Image Segmentation. arXiv. 2018; DOI: 10.48550/arxiv.1809.10486.
5. Wasserthal J, Breit H-C, Meyer MT, Pradella M, Hinck D, Sauter AW, et al. Totalsegmentator: robust segmentation of 104 anatomic structures in CT images. Radiol Artif Intell. 2023 Sep;5(5):e230024. DOI: 10.1148/ryai.230024. PMID: 37795137. PMCID: PMC10546353.
6. Pickhardt PJ, Summers RM, Garrett JW, Krishnaraj A, Agarwal S, Dreyer KJ, et al. Opportunistic screening: radiology scientific expert panel. Radiology. 2023 Jun;307(5):e222044. DOI: 10.1148/radiol.222044. PMID: 37219444. PMCID: PMC10315516.
7. Fischl B, Salat DH, van der Kouwe AJW, Makris N, Ségonne F, Quinn BT, et al. Sequence-independent segmentation of magnetic resonance images. Neuroimage. 2004;23 Suppl 1:S69-84. DOI: 10.1016/j.neuroimage.2004.07.016. PMID: 15501102.
8. Fedorov A, Longabaugh WJR, Pot D, Clunie DA, Pieper SD, Gibbs DL, et al. National cancer institute imaging data commons: toward transparency, reproducibility, and scalability in imaging artificial intelligence. Radiographics. 2023 Dec;43(12):e230180. DOI: 10.1148/rg.230180. PMID: 37999984. PMCID: PMC10716669.
9. Ji Y, Bai H, Yang J, Ge C, Zhu Y, Zhang R, et al. AMOS: A Large-Scale Abdominal Multi-Organ Benchmark for Versatile Medical Image Segmentation. arXiv. 2022; DOI: 10.48550/arxiv.2206.08023.
10. Kavur AE, Gezer NS, Barış M, Aslan S, Conze P-H, Groza V, et al. CHAOS Challenge - combined (CT-MR) healthy abdominal organ segmentation. Med Image Anal. 2021 Apr;69:101950. DOI: 10.1016/j.media.2020.101950. PMID: 33421920.
11. Anastasopoulos C, Reisert M, Kellner E. "Nora Imaging": A Web-Based Platform for Medical Imaging. Neuropediatrics. 2017 Apr 26;48(S 01):S1–45. DOI: 10.1055/s-0037-1602977.
12. Hoopes A, Mora JS, Dalca AV, Fischl B, Hoffmann M. SynthStrip: skull-stripping for any brain image. Neuroimage. 2022 Oct 15;260:119474. DOI: 10.1016/j.neuroimage.2022.119474. PMID: 35842095. PMCID: PMC9465771.
13. Häntze H, Xu L, Dorfner FJ, Donle L, Truhn D, Aerts H, et al. MRSegmentator: Robust Multi-Modality Segmentation of 40 Classes in MRI and CT Sequences. arXiv. 2024; DOI: 10.48550/arxiv.2405.06463.
14. Maier-Hein L, Reinke A, Godau P, Tizabi MD, Buettner F, Christodoulou E, et al. Metrics reloaded: recommendations for image analysis validation. Nat Methods. 2024 Feb 12;21(2):195–212. DOI: 10.1038/s41592-023-02151-z. PMID: 38347141. PMCID: PMC11182665.
15. Zhou A, Liu Z, Tieu A, Patel N, Sun S, Yang A, et al. MRAnnotator: A Multi-Anatomy Deep Learning Model for MRI Segmentation. arXiv. 2024; DOI: 10.48550/arxiv.2402.01031.
16. Zhuang Y, Mathai TS, Mukherjee P, Khoury B, Kim B, Hou B, et al. MRISegmentator-Abdomen: A Fully Automated Multi-Organ and Structure Segmentation Tool for T1-weighted Abdominal MRI. arXiv. 2024; DOI: 10.48550/arxiv.2405.05944.

17. Ma J, He Y, Li F, Han L, You C, Wang B. Segment anything in medical images. Nat Commun. 2024 Jan 22;15(1):654. DOI: 10.1038/s41467-024-44824-z. PMCID: PMC10803759.

18. Zhu J, Qi Y, Wu J. Medical SAM 2: Segment medical images as video via Segment Anything Model 2. arXiv. 2024; DOI: 10.48550/arxiv.2408.00874.

19. Isensee F, Wald T, Ulrich C, Baumgartner M, Roy S, Maier-Hein K, et al. nnU-Net Revisited: A Call for Rigorous Validation in 3D Medical Image Segmentation. arXiv. 2024; DOI: 10.48550/arxiv.2404.09556.

# Supplemental Materials

## S1. List of sources of external IDC datasets

1. Radiology Data from The Cancer Genome Atlas Liver Hepatocellular Carcinoma [TCGA-LIHC] collection (1)
2. A radiomics model from joint FDG-PET and MRI texture features for the prediction of lung metastases in soft-tissue sarcomas of the extremities (2)
3. Radiology Data from the Clinical Proteomic Tumor Analysis Consortium Clear Cell Renal Cell Carcinoma [CPTAC-CCRCC] collection (3)
4. Radiology Data from The Cancer Genome Atlas Urothelial Bladder Carcinoma [TCGA-BLCA] collection (4)
5. Radiology Data from The Cancer Genome Atlas Cervical Kidney renal papillary cell carcinoma [KIRP] collection (5)
6. Radiology Data from The Cancer Genome Atlas Kidney Renal Clear Cell Carcinoma [TCGA-KIRC] collection (6)
7. Data from the ACRIN 6668 Trial NSCLC-FDG-PET (7)
8. The Clinical Proteomic Tumor Analysis Consortium Sarcomas Collection (CPTAC-SAR) (8)
9. Annotations for The Clinical Proteomic Tumor Analysis Consortium Clear Cell Renal Cell Carcinoma Collection (CPTAC-CCRCC-Tumor-Annotations) (9)
10. Radiology Data from the Clinical Proteomic Tumor Analysis Consortium Pancreatic Ductal Adenocarcinoma [CPTAC-PDA] Collection (10)
11. Annotations for The Clinical Proteomic Tumor Analysis Consortium Pancreatic Ductal Adenocarcinoma Collection (CPTAC-PDA-Tumor-Annotations) (11)
12. Radiology Data from The Cancer Genome Atlas Kidney Chromophobe [TCGA-KICH] collection (12)
13. Radiology Data from The Cancer Genome Atlas Sarcoma [TCGA-SARC] collection (13)
14. Radiology Data from the Clinical Proteomic Tumor Analysis Consortium Lung Adenocarcinoma [CPTAC-LUAD] collection (14)
15. Radiology Data from the Clinical Proteomic Tumor Analysis Consortium Cutaneous Melanoma [CPTAC-CM] collection (15)
16. Stony Brook University COVID-19 Positive Cases (16)
17. Cancer Moonshot Biobank - Lung Cancer Collection (CMB-LCA) (17)
18. Cancer Moonshot Biobank - Multiple Myeloma Collection (CMB-MML) (18)
19. Cancer Moonshot Biobank - Colorectal Cancer Collection (CMB-CRC) (19)
20. A DICOM dataset for evaluation of medical image de-identification (Pseudo-PHI-DICOM-Data (20)
21. Image segmentations produced by BAMF under the AIMI Annotations initiative (21)
22. National cancer institute imaging data commons: toward transparency, reproducibility, and scalability in imaging artificial intelligence. (22)

# S2. List of all segmented structures

Spleen, kidney right, kidney left, gallbladder, liver, stomach, pancreas, adrenal gland right, adrenal gland left, lung left, lung right, esophagus, small bowel, duodenum, colon, urinary bladder, prostate, sacrum, vertebrae, intervertebral discs, spinal cord, heart, aorta, inferior vena cava, portal vein and splenic vein, iliac artery left, iliac artery right, iliac vena left, iliac vena right, humerus left, humerus right, scapula left, scapula right, clavicula left, clavicula right, femur left, femur right, hip left, hip right, gluteus maximus left, gluteus maximus right, gluteus medius left, gluteus medius right, gluteus minimus left, gluteus minimus right, back muscle left, back muscle right, iliopsoas left, iliopsoas right, brain, quadriceps femoris left, quadriceps femoris right, thigh medial compartment left, thigh medial compartment right, thigh posterior compartment left, thigh posterior compartment right, sartorius left, sartorius right, deltoid, supraspinatus, infraspinatus, subscapularis, coracobrachial, trapezius, pectoralis minor, serratus anterior, teres major, triceps brachii, patella, tibia, fibula, tarsal, metatarsal, phalanges feet, ulna, radius, face, subcutaneous adipose tissue, skeletal muscle, visceral adipose tissue.

# S3. List of all pretrained models which were used during the data annotation

1. **Name**: synthstrip (https://surfer.nmr.mgh.harvard.edu/docs/synthstrip/)
   **Structures:** brain
2. **Name**: https://doi.org/10.1186/s12880-023-01056-9
   **Structures:** gluteus left, gluteus right, quadriceps femoris left, quadriceps femoris right
3. **Name**: MRSegmentator (https://github.com/hhaentze/MRSegmentator)
   **Structures:** gallbladder, stomach, adrenal gland right, adrenal gland left, small bowel, duodenum, colon, portal vein and splenic vein, iliac artery left, iliac artery right, iliac vena left, iliac vena right
4. **Name:** https://doi.org/10.1016/j.ebiom.2024.105467
   **Structures:** subcutaneous adipose tissue, visceral adipose tissue, skeletal muscle

# S4. Training details

The following nnU-Net settings were applied: Only the model 3d_fullres was used. Mirroring was removed from the data augmentation pipeline because otherwise, the model was not able to properly distinguish between left and right anatomical structures. Normalization was set to "MR" for all our experiments even when only training on CT images. The number of training epochs was set to 2000 epochs because of the large dataset size. No ensembling was applied to keep runtime low.
Training one nnU-Net model for all 80 structures at the same time results in very high memory consumption. Thus, we split the 80 structures into six parts each with a lower number of structures . For each part, one smaller nnU-Net was trained.
For the TotalSegmentator MRI-3 it was possible to combine all 50 main structures into one model without unreasonable memory requirements.

# S5. List of segmented structures for comparison to TotalSegmentator MRI-3 and MRSegmentator and AMOS model

**TotalSegmentator MRI-3 (50 structures)**: spleen, kidney right, kidney left, gallbladder, liver, stomach, pancreas, adrenal gland right, adrenal gland left, lung left, lung right, esophagus, small bowel, duodenum, colon, urinary bladder, prostate, sacrum, vertebrae, intervertebral discs, spinal cord, heart, aorta, inferior vena cava, portal vein and splenic vein, iliac artery left, iliac artery right, iliac vena left, iliac vena right, humerus left, humerus right, scapula left, scapula right, clavicula left, clavicula right, femur left, femur right, hip left, hip right, gluteus maximus left, gluteus maximus right, gluteus medius left, gluteus medius right, gluteus minimus left, gluteus minimus right, back muscle left, back muscle right, iliopsoas left, iliopsoas right, brain.

**MRSegmentator (40 structures)**: spleen, kidney right, kidney left, gallbladder, liver, stomach, pancreas, adrenal gland right, adrenal gland left, lung left, lung right, heart, aorta, inferior vena cava, portal vein and splenic vein, iliac artery left, iliac artery right, iliac vena left, iliac vena right, esophagus, small bowel, duodenum, colon, urinary bladder, vertebrae, sacrum, hip left, hip right, femur left, femur right, back muscle left, back muscle right, iliopsoas left, iliopsoas right, gluteus maximus left, gluteus maximus right, gluteus medius left, gluteus medius right, gluteus minimus left, gluteus minimus right.

**AMOS (13 structures)**: spleen, kidney right, kidney left, gallbladder, liver, stomach, pancreas, adrenal gland right, adrenal gland left, esophagus, duodenum, aorta, inferior vena cava.

# S6. Supplemental Table

**Supplemental Table.** The table shows the cut-off volume in ml with p-values for every body structure regarding volume comparing the age quartiles. Group comparison was performed using the Kruskal-Wallis test. Significance was reported if the p-value was below 0.0001 due to Bonferroni correction. If a segmentation of a specific body region of a patient was below the lower cut-off, the segmentation was excluded.

| Region | Lower Cut Off Volume in ml | Volume p-value | Volume significant |
|---|---|---|---|
| spleen | 40 | < 0.0001 | Yes |
| right kidney | 50 | < 0.0001 | Yes |
| left kidney | 30 | < 0.0001 | Yes |
| liver | 100 | < 0.0001 | Yes |
| right adrenal gland | 1 | < 0.0001 | Yes |
| left adrenal gland | 1 | < 0.0001 | Yes |

# S7. Supplemental Figures

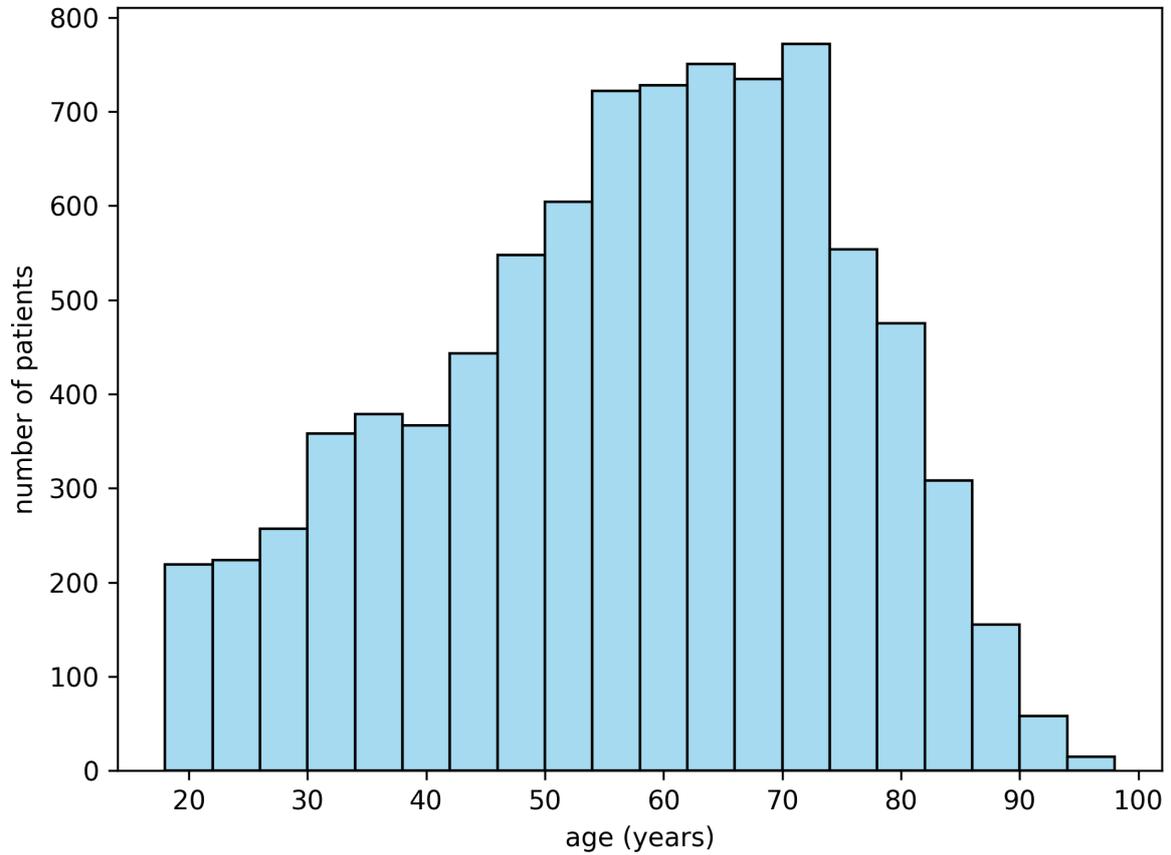

Figure S1. Age distribution of the aging study dataset.

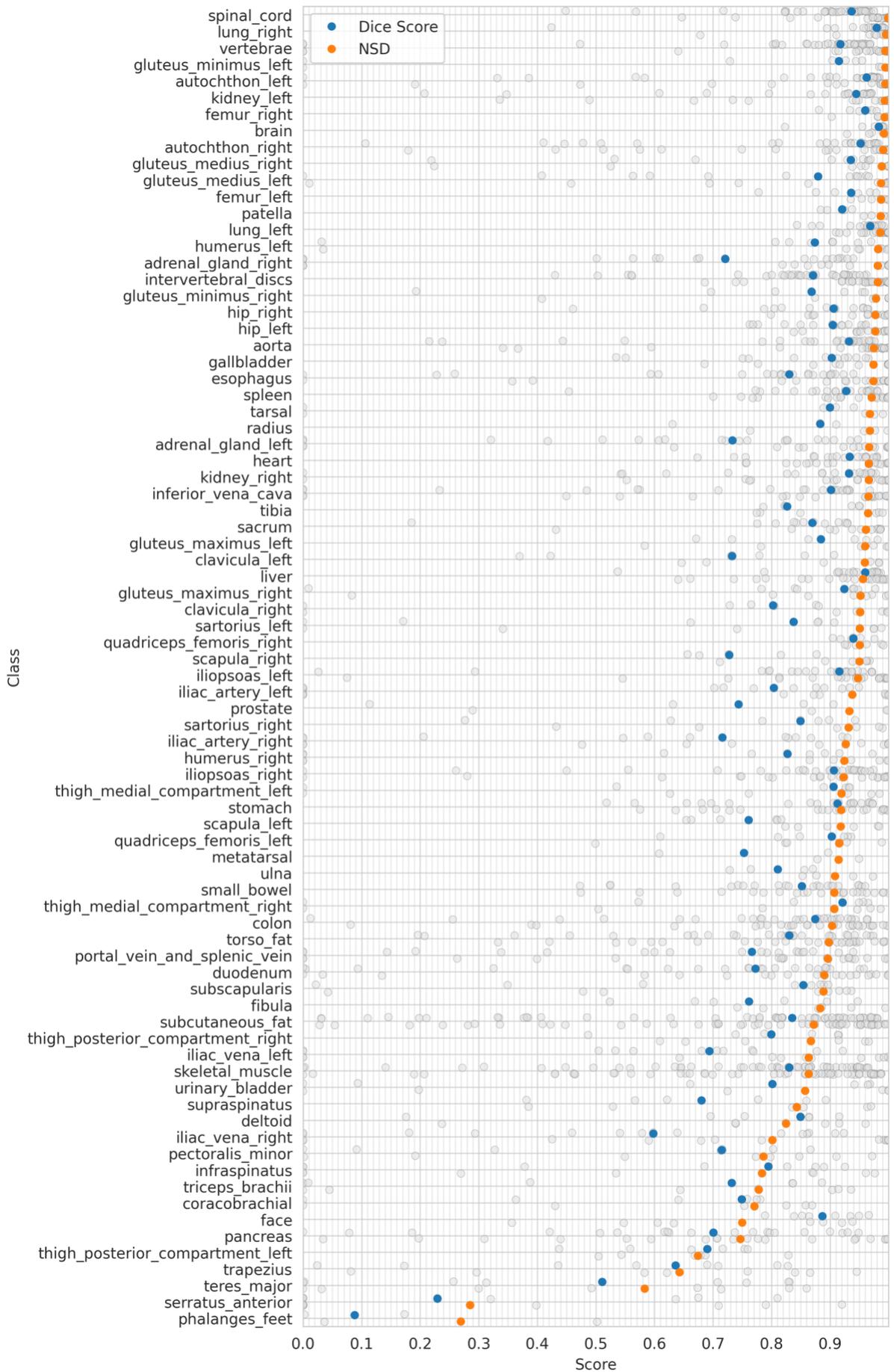

Figure S2. Overview of results of our model for each anatomical structure, sorted by Dice score and Normalised Surface Distance (NSD). The blue dots show the median of the Dice score, and the orange dots the median of the NSD. The grey dots show the results for each subject.

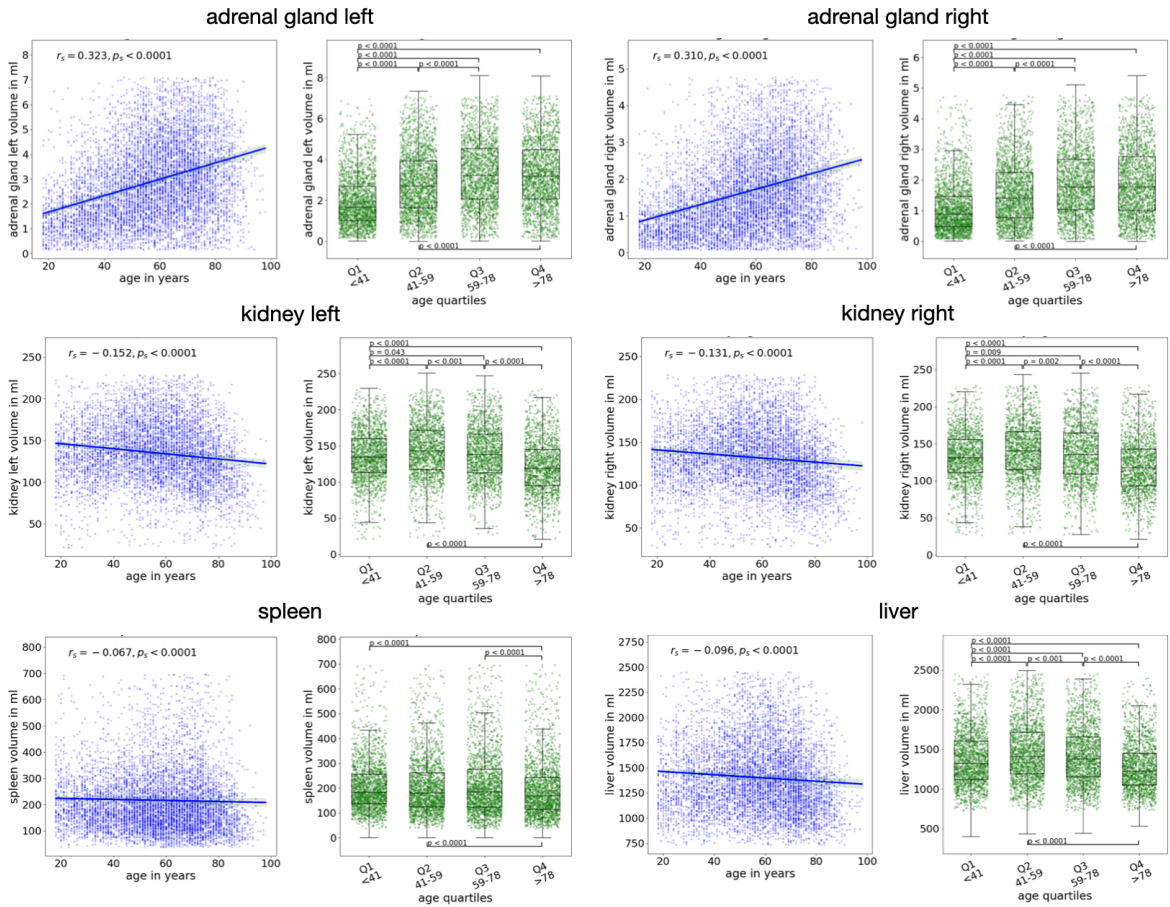

Figure S3. Scatterplot with correlation (blue) and boxplot of age quartiles (green) showing changes of organ volume over the lifetime for six different abdominal organs.

# S8. Supplemental References

1. Erickson BJ, Kirk S, Lee Y, Bathe O, Kearns M, Gerdes C, et al. Radiology Data from The Cancer Genome Atlas Liver Hepatocellular Carcinoma [TCGA-LIHC] collection. The Cancer Imaging Archive. 2016; DOI: 10.7937/k9/tcia.2016.immqw8uq.
2. Vallières M, Freeman CR, Skamene SR, El Naqa I. A radiomics model from joint FDG-PET and MRI texture features for the prediction of lung metastases in soft-tissue sarcomas of the extremities. Phys Med Biol. 2015 Jul 21;60(14):5471–96. DOI: 10.1088/0031-9155/60/14/5471. PMID: 26119045.
3. National Cancer Institute Clinical Proteomic Tumor Analysis Consortium (CPTAC). Radiology Data from the Clinical Proteomic Tumor Analysis Consortium Clear Cell Renal Cell Carcinoma [CPTAC-CCRCC] collection. The Cancer Imaging Archive. 2018; DOI: 10.7937/k9/tcia.2018.oblamn27.
4. Kirk S, Lee Y, Lucchesi FR, Aredes ND, Gruszauskas N, Catto J, et al. Radiology Data from The Cancer Genome Atlas Urothelial Bladder Carcinoma [TCGA-BLCA] collection. The Cancer Imaging Archive. 2016; DOI: 10.7937/k9/tcia.2016.8lng8xdr.
5. Linehan M, Gautam R, Kirk S, Lee Y, Roche C, Bonaccio E, et al. Radiology Data from The Cancer Genome Atlas Cervical Kidney renal papillary cell carcinoma [KIRP] collection. The Cancer Imaging Archive. 2016; DOI: 10.7937/k9/tcia.2016.acwogbef.
6. Akin O, Elnajjar P, Heller M, Jarosz R, Erickson BJ, Kirk S, et al. Radiology Data from The Cancer Genome Atlas Kidney Renal Clear Cell Carcinoma [TCGA-KIRC] collection. The Cancer Imaging Archive. 2016; DOI: 10.7937/k9/tcia.2016.v6pbvtdr.
7. Kinahan P, Muzi M, Bialecki B, Herman B, Coombs L. Data from the ACRIN 6668 Trial NSCLC-FDG-PET. The Cancer Imaging Archive. 2019; DOI: 10.7937/tcia.2019.30ilqfcl.
8. National Cancer Institute Clinical Proteomic Tumor Analysis Consortium (CPTAC). The Clinical Proteomic Tumor Analysis Consortium Sarcomas Collection (CPTAC-SAR). The Cancer Imaging Archive. 2019; DOI: 10.7937/tcia.2019.9bt23r95.
9. Rozenfeld M, Jordan P. Annotations for The Clinical Proteomic Tumor Analysis Consortium Clear Cell Renal Cell Carcinoma Collection (CPTAC-CCRCC-Tumor-Annotations). The Cancer Imaging Archive. 2023; DOI: 10.7937/skq4-qx48.
10. National Cancer Institute Clinical Proteomic Tumor Analysis Consortium (CPTAC). Radiology Data from the Clinical Proteomic Tumor Analysis Consortium Pancreatic Ductal Adenocarcinoma [CPTAC-PDA] Collection. The Cancer Imaging Archive. 2018; DOI: 10.7937/k9/tcia.2018.sc20fo18.
11. Rozenfeld M, Jordan P. Annotations for The Clinical Proteomic Tumor Analysis Consortium Pancreatic Ductal Adenocarcinoma Collection (CPTAC-PDA-Tumor-Annotations). The Cancer Imaging Archive. 2023; DOI: 10.7937/bw9v-bx61.
12. Linehan MW, Gautam R, Sadow CA, Levine S. Radiology Data from The Cancer Genome Atlas Kidney Chromophobe [TCGA-KICH] collection. The Cancer Imaging Archive. 2016; DOI: 10.7937/k9/tcia.2016.yu3rbczn.
13. Roche C, Bonaccio E, Filippini J. Radiology Data from The Cancer Genome Atlas Sarcoma [TCGA-SARC] collection. The Cancer Imaging Archive. 2016; DOI: 10.7937/k9/tcia.2016.cx6ylsux.
14. National Cancer Institute Clinical Proteomic Tumor Analysis Consortium (CPTAC). Radiology Data from the Clinical Proteomic Tumor Analysis Consortium Lung Adenocarcinoma [CPTAC-LUAD] collection. The Cancer Imaging Archive. 2018; DOI: 10.7937/k9/tcia.2018.pat12tbs.

15.	National Cancer Institute Clinical Proteomic Tumor Analysis Consortium (CPTAC). Radiology Data from the Clinical Proteomic Tumor Analysis Consortium Cutaneous Melanoma [CPTAC-CM] collection. The Cancer Imaging Archive. 2018; DOI: 10.7937/k9/tcia.2018.odu24gze.
16.	Saltz J, Saltz M, Prasanna P, Moffitt R, Hajagos J, Bremer E, et al. Stony Brook University COVID-19 Positive Cases. The Cancer Imaging Archive. 2021; DOI: 10.7937/tcia.bbag-2923.
17.	Biobank CM. Cancer Moonshot Biobank - Lung Cancer Collection (CMB-LCA). The Cancer Imaging Archive. 2022; DOI: 10.7937/3cx3-s132.
18.	Biobank CM. Cancer Moonshot Biobank - Multiple Myeloma Collection (CMB-MML). The Cancer Imaging Archive. 2022; DOI: 10.7937/szkb-sw39.
19.	Biobank CM. Cancer Moonshot Biobank - Colorectal Cancer Collection (CMB-CRC). The Cancer Imaging Archive. 2022; DOI: 10.7937/djg7-gz87.
20.	Rutherford M, Mun SK, Levine B, Bennett W, Smith K, Farmer P, et al. A DICOM dataset for evaluation of medical image de-identification. Sci Data. 2021 Jul 16;8(1):183. DOI: 10.1038/s41597-021-00967-y. PMID: 34272388. PMCID: PMC8285420.
21.	Van Oss J, Murugesan GK, McCrumb D, Soni R. Image segmentations produced by BAMF under the AIMI Annotations initiative. Zenodo. 2023; DOI: 10.5281/zenodo.10081112.
22.	Fedorov A, Longabaugh WJR, Pot D, Clunie DA, Pieper SD, Gibbs DL, et al. National cancer institute imaging data commons: toward transparency, reproducibility, and scalability in imaging artificial intelligence. Radiographics. 2023 Dec;43(12):e230180. DOI: 10.1148/rg.230180. PMID: 37999984. PMCID: PMC10716669.